\documentstyle[aas2pp4,psfig]{article}

\newcommand\mdot{$\dot M$}

\newcommand\about{$\sim$}

\def\hide#1{}
\def\mk{van der Klis, M.}
\def\lw{Lewin, W.H.G.}
\def\vpj{van Paradijs, J.}
\righthead{Kilohertz QPO in Sco X-1}
\slugcomment{1997 March 5 -- To appear in ApJ Letters} 
\begin{document}

\title{\noindent{\normalsize\tt 1997 March 5 -- To appear in ApJ Letters \vskip 24pt}
Kilohertz QPO Peak Separation Is Not Constant in Scorpius X-1}

\author{Michiel van der Klis and Rudy\,A.D. Wijnands}
\affil{Astronomical Institute ``Anton Pannekoek'', University of
Amsterdam \\and Center for High-Energy Astrophysics, Amsterdam, The
Netherlands; \\michiel@astro.uva.nl,rudy@astro.uva.nl}

\author{Keith Horne}
\affil{School of Physics and Astronomy, North Haugh, St. Andrews KY16
9SS, Scotland, UK; \\kdh1@st-andrews.ac.uk}

\author{Wan Chen}
\affil{NASA/GSFC Lab. for High-Energy Astrophysics, 
Goddard Space Flight Center, USA; \\chen@rosserv.gsfc.nasa.gov}

\begin{abstract}
We report on a series of twenty \about10$^5$\,c/s, 0.125\,msec
time-resolution RXTE observations of the Z source and low-mass X-ray
binary Scorpius\,X-1. Twin kilohertz quasi-periodic oscillation (QPO)
peaks are obvious in nearly all observations. We find that the
peak separation is not constant, as expected in some beat-frequency
models, but instead varies from \about310 to \about230\,Hz when the
centroid frequency of the higher-frequency peak varies from \about875
to \about1085\,Hz. We detect none of the additional QPO peaks at
higher frequencies predicted in the photon bubble model (PBM), with
best-case upper limits on the peaks' power ratio of 0.025. We do
detect, simultaneously with the kHz QPO, additional QPO peaks near 45
and 90\,Hz whose frequency increases with mass accretion rate. We
interpret these as first and second harmonics of the so-called
horizontal-branch oscillations well known from other Z sources and
usually interpreted in terms of the magnetospheric beat-frequency
model (BFM). We conclude that the magnetospheric BFM and the PBM are
now unlikely to explain the kHz QPO in Sco\,X-1. In order to succeed
in doing so, {\it any} BFM involving the neutron star spin (unseen in
Sco\,X-1) will have to postulate at least one {\it additional} unseen
frequency, beating with the spin to produce one of the kHz peaks.

\end{abstract}


\section{Introduction}

Kilohertz quasi-periodic oscillations (QPO) have now been observed in
eleven low-mass X-ray binaries, the Z sources Sco\,X-1 (van der Klis
et al. 1996a; hereafter Paper\,1), GX\,5$-$1 (van der Klis et
al. 1996b) and GX\,17+2 (van der Klis et al. 1997), the atoll sources
(see Hasinger and van der Klis 1989) 4U\,1728$-$34 (Strohmayer et
al. 1996a), 4U\,1608$-$52 (Berger et al. 1996), 4U\,1636$-$53 (Zhang
et al. 1996), 4U\,0614+09 (Ford et al. 1997), 4U\,1735$-$44 (Wijnands
et al. 1996a) and 4U\,1820$-$30 (Smale, Zhang and White 1996), and in
KS\,1731$-$260 (Morgan and Smith 1996) and a source near the galactic
center, perhaps \break MXB\,1743$-$29 (Strohmayer et al. 1996b). In most of
these sources, the QPO frequency has been observed to increase with
accretion rate \mdot; frequencies are in the range 325--1193\,Hz and
relative peak widths vary between 0.11\% and 10\%. Most often {\it
double} peaks are observed, with a separation in the range
250--360\,Hz. In 4U\,1728$-$34 (Strohmayer et al. 1996a), during X-ray
bursts, a third peak is seen near a frequency of 360\,Hz, compatible
with the separation of the two kHz peaks, which remains constant as
the peaks move up and down in frequency. Three other cases of three
similarly commensurate frequencies have been reported (4U\,0614+09;
Ford et al. 1997, 4U\,1636$-$53; Zhang et al. 1997, and
KS\,1731$-$260; Wijnands and van der Klis 1997). This strongly
suggests a beat-frequency interpretation, with the \about360\,Hz peak
in 4U\,1728$-$34 at the neutron star spin frequency, the
higher-frequency (hereafter ``upper'') kHz peak at the Kepler
frequency corresponding to some preferred orbital radius around the
neutron star, and the lower-frequency (hereafter ``lower'') kHz peak
at the difference frequency between these two. Strohmayer et
al. (1996a) suggested that this preferred radius is the magnetospheric
radius. Miller, Lamb and Psaltis (1996) proposed it is the sonic
radius.

Although ways out can always be found, this class of models naturally
predicts the peak separation to be constant. In this paper we present
data that show conclusively that in Sco\,X-1 the peak separation
varies systematically. A brief announcement of this result already
appeared in van der Klis et al. (1996c). We also present evidence for
the presence of horizontal branch oscillations (HBO; see van der Klis
1995 for a recent summary of Z-source characteristics) in Sco\,X-1
near 45\,Hz with a harmonic near 90\,Hz. This is the first time that
HBO have been positively identified in Sco\,X-1. HBO are usually
interpreted in terms of the magnetospheric beat frequency model (Alpar
and Shaham 1985, Lamb et al. 1985), precluding the application of this
model to the kHz QPO that occur at the same time.

\section{Observations and analysis}

We observed Sco\,X-1 with the RXTE PCA (Bradt, Rothschild and Swank
1993) 20 times during May 24--28, 1996. Each observation consisted of
2--4 continuous data intervals of 1--3\,ks each. Single- and
double-event data (Paper\,1) were recorded in parallel and combined
off-line to enhance sensitivity. A time resolution of 1/8192\,s
(\about0.125\,ms) was used throughout.

During these observations various offset angles were used, and not all
5 detectors were always on. For these reasons, the expected Z track in
the X-ray color-color diagram can not now be recovered; this awaits
better understanding of the spectral calibration of the PCA at high
count rates and off-axis source positions. Raw count rates varied
between 60 and 130\,kc/s (2--60\,keV).

\begin{figure}[htb]
\begin{center}
\begin{tabular}{c}
\psfig{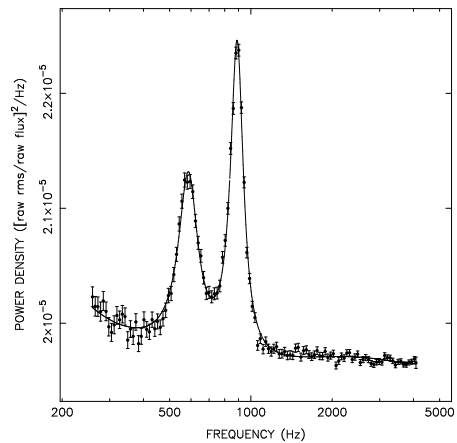}
\end{tabular}
\caption{Power spectrum from \about10\,ks of data 
showing double kHz QPO peaks, with best fit superimposed. Note the
absence of additional peaks. The sloping continuum above 1\,kHz is
instrumental (\S2). \label{fig1}}
\end{center}
\end{figure}

We calculated power spectra of all 0.125\,ms data using 16-s data
segments and calculated one average spectrum for each continuous data
interval. For measuring the properties of the kHz QPO we fitted the
256--4096\,Hz power spectra (Fig.\,1) with a function consisting of a
constant, two Lorentzian peaks and either a broad sinc or a broad
sinusoid to represent the deadtime-modified Poisson noise, depending
on the Very Large Event window setting (Zhang et al. 1995, Zhang
1995). The PCA deadtime process at 10$^5$\,c/s is not, as yet,
sufficiently well understood to accurately predict this Poisson
component. We can not, therefore, report on the properties of any
intrinsic broad noise components in the kHz range.

For measuring the 45\,Hz QPO and its harmonic we fitted the
16--256\,Hz power spectra with a broad Lorentzian centered near zero
frequency to represent the continuum, and one or two Lorentzian peaks
to model the QPO. The conversion of the power in the QPO peaks to
fractional rms amplitude depends on the derivative of the deadtime
transmission function with respect to count rate (van der Klis 1989),
which we do not know. The deadtime is expected to suppress the QPO
amplitude more than the total count rate. Our reported raw (i.e.,
uncorrected for deadtime) fractional rms amplitudes are therefore
lower limits to the true values. These could be several times as
large.

\section{Results}

Kilohertz QPO were detected in all observations. The peaks (Fig.\,1)
are very significant, with raw rms values of up to 2.5\%, and the
spectra are well-fit by the fit function described in \S2. Fig.\,2
illustrates the changes in power-spectral shape as a function of
inferred \mdot.  Notice the increase in frequency and decrease in
power of the two kHz QPO peaks, the emergence of the normal-branch
oscillations (NBO) near 6\,Hz apparently {\it from} the low-frequency
noise (LFN), and the complicated variations in strength and shape of
the 45 and 90\,Hz peaks with \mdot\ (increasing upwards). As in
Paper\,1, the frequency of the NBO is correlated to that of the kHz QPO.

\begin{figure}[htbp]
\begin{center}
\begin{tabular}{c}
\psfig{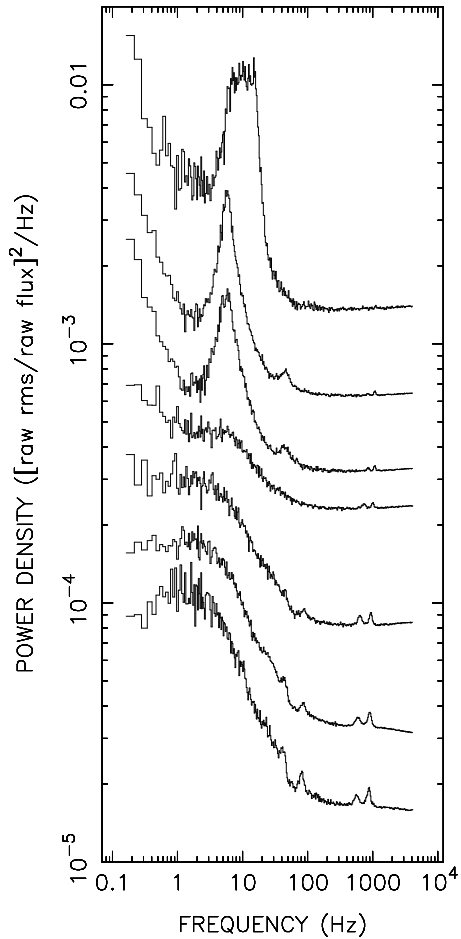}
\end{tabular}
\caption{Representative \about3\,ks power spectra sorted 
according to inferred \mdot\ (increasing upwards) and shifted up by
factors of 1, 2, 5, 10, 20, 40 and 80, respectively, for clarity. The
frequency of the upper kHz peak increases with \mdot\ from 872 to
1115\,Hz, that of the lower one from 565 to 890\,Hz. The large width
of the 10-Hz peak in the top trace is due to peak motion. The sloping
continua in the kHz range are instrumental (\S2). \label{fig2}}
\end{center}
\end{figure}

As we can not estimate \mdot\ from the X-ray color-color diagram, we
plot in Fig.\,3 the results of our fits vs. the centroid frequency
$\nu_u$ of the upper peak. $\nu_u$ increases monotonically with
position along the Z track, and, by inference, \mdot\
(Paper\,1). Inspection of the variations in the properties of the
NBO, whose relation to \mdot\ is known, confirms this.

\begin{figure}[hbtp]
\begin{center}
\begin{tabular}{c}
\psfig{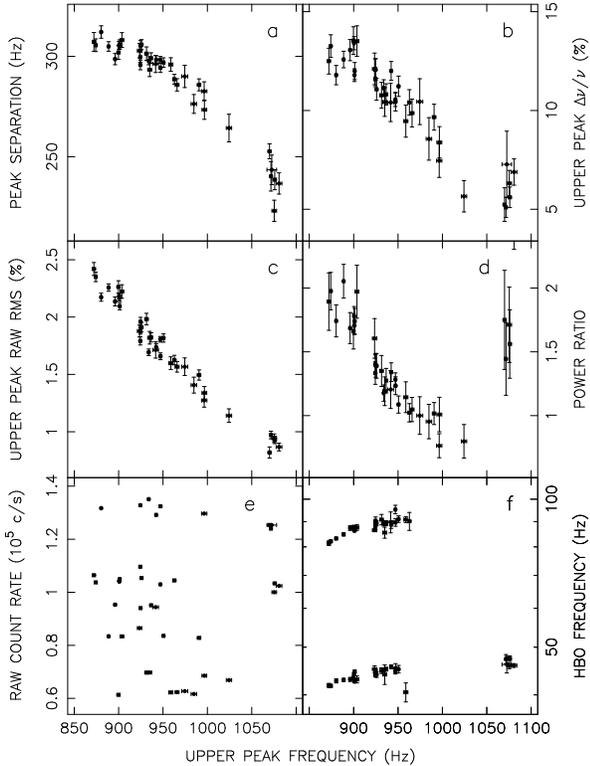}
\end{tabular}
\caption{Kilohertz peak separation ({\it a}), relative 
width (full width at half maximum divided by centroid frequency,
expressed in percent; {\it b}) and raw fractional rms amplitude ({\it
c}) of the highest frequency (``upper'') peak, ratio of the power in
the upper to that in the lower peak ({\it d}), raw count rate ({\it
e}) and horizontal branch oscillation first and second harmonics'
frequency ({\it f}) as a function of the frequency of the upper kHz
peak. One data point at 1075\,Hz with a value near 3 is off-scale in
({\it d}). Eight low signal-to-noise data points were not plotted. The
gap between 1000 and 1075\,Hz is due to a lack of good
data.\label{fig3}}
\end{center}
\end{figure}

Fig.\,3{\it a} shows the variation of the kHz peak separation. There
is a strong decrease with $\nu_u$: when their frequencies increase,
the two peaks systematically move closer together. The relation of
peak separation to $\nu_u$ appears to be non-linear. Fig.\,3{\it b}
shows that the higher-frequency QPO become more coherent as their
frequency increases. The same trend is present, less pronounced, in
the lower-frequency QPO. The ratio of upper to lower kHz-peak width
(not shown) is roughly constant at 1. Fig.\,3{\it c} shows how the raw
fractional rms amplitude of the upper peak falls with \mdot. Again,
the lower peak shows a similar trend, but less pronounced. Although
the raw rms values are deadtime- and therefore count-rate dependent
(\S2), the changes in raw count rate are random (Fig.\,3{\it e}), so
that the trend in rms must be intrinsic to the source. Fig.\,3{\it d}
shows the kHz peaks' power ratio (which is essentially free of
deadtime effects, as any time scale dependence in these effects is
expected to become appreciable only for time scales near the deadtime,
\about10\,$\mu$sec; Zhang 1995). The power ratio varies 
non-monotonically as a function of $\nu_u$, dropping by more than a
factor of 2 between 875 and 1000\,Hz, then increasing again.

We detect no other kHz QPO peaks. When $\nu_u$ $<$1\,kHz and the
detected peaks are relatively strong, any other kHz peaks of similar
width are typically less than 0.025--0.2 times the power in the
detected peaks (95\% confidence), depending on data selection. For
$\nu_u$$>$1\,kHz these limits worsen as the detected peaks get weaker.

Fig.\,3{\it f}, finally, shows the frequencies of the \about45 and
\about90\,Hz peaks. For $\nu_u$$<$960\,Hz the 45\,Hz peak has a width 
of \about8\,Hz and the 90\,Hz peak of 20--40\,Hz. Both peaks drop
rapidly in raw rms amplitude as $\nu_u$ increases, from \about1\% at
$\nu_u$\about850 to below the detection limit of \about0.4\% between
960 and 1000\,Hz. There are no good data between 1000 and 1075\,Hz. At
1075\,Hz the \about45\,Hz peak reappears at 1\% rms, now much broader
(20--40\,Hz; see also Fig.\,2).

\section{Discussion}

Our data place severe constraints on any model for kHz QPO so far
discussed.  We note, that the twin kHz QPO in Sco\,X-1 and those
in the atoll sources are likely to be the same phenomenon. The
frequencies, their dependence on \mdot, the coherencies, the peak
separations and the fact that there are {\it two} peaks, one of
which sometimes becomes undetectable at extreme \mdot, are too similar
to be attributed to just coincidence. Even the amplitude of the kHz
QPO in Sco\,X-1 is in the range of that seen in the atoll sources
(although in some of them much higher values have been seen). This
implies, then, that the variable peak separation we detect in Sco\,X-1
must be explained within the same model as the properties of the
twin peaks in atoll sources.

The 45 and 90\,Hz QPO we reported here are nearly certainly the first
and second harmonic of the horizontal branch QPO (HBO), usually seen
in Z sources in the horizontal and upper normal branches (see van der
Klis 1995). This was already tentatively suggested for the broad
45\,Hz peak seen when $\nu_u$$>$1075\,Hz in Paper\,1. The similarities
with HBO in other Z sources include the frequency, its increase (but
see Wij\-nands et al. 1996), and the decrease in rms, with \mdot, and
the presence of a 2nd harmonic.  This constitutes the first positive
identification of HBO in Sco\,X-1.

The morphology of the spectra in Fig.\,2 seems to suggest that the
well-known, slightly peaked broad noise component below 20\,Hz usually
called \break low-frequency noise (LFN) ``peaks up'' into the, also
well-known, 6\,Hz normal branch QPO (NBO) when \mdot\
increases. Appearances may deceive. One way to check whether this
suggested relation between NBO and LFN is real would be to study the
photon energy dependence of the amplitude and phase of the
variability, which in NBO can be quite characteristic (Mitsuda and
Dotani 1989). There is a small shoulder in our spectra below the
45\,Hz peak (see, e.g., Fig.\,2, second spectrum from below) that can
perhaps be identified as the true signature of the noise component
that is expected to accompany HBO (cf. Lamb et al. 1985).

The photon bubble model (Klein et al. 1996), and also some neutron
star vibration models, predict several kHz QPO peaks at frequencies
above those of the two detected ones, and of similar strength as
these.  The fact that we observe just these two kHz peaks, with good
upper limits on any additional ones, is a strong argument against
these models.

If the kHz QPO are due to a millisecond X-ray pulsar whose pulsations
we see (Doppler-shifted) reflected off inhomogeneities in the Fortner
et al. (1989) radial flow (Paper\,1), then one would expect the QPO to
become weaker at low \mdot, when the radial flow subsides; instead we
find that the kHz QPO become much stronger at low \mdot. In a variant
on this model, suppose that two relativistic jets are emerging in
opposite directions from near the neutron star, and that we are seeing
the signal from a central millisecond X-ray pulsar, not directly, but
reflected off inhomogeneities in the jets. Because there are two jets,
there are two QPO peaks, at
$\nu_\mp=\nu_{pulse}(1-v/c)/(1\pm(v/c)\cos\theta)$, where $v$ is the
jets' speed, $\theta$ their angle with the line of sight, and
$\nu_{pulse}$ the unseen pulse frequency (van der Klis 1996). Such a
model fits very well to the non-linear relation plotted in Fig.\,3{\it
a}, with $\nu_{pulse}$=1370\,Hz (which could be twice the spin rate of
the neutron star) and $\theta$=61$^\circ$, if $v/c$ decreases from
0.48 to 0.26 with increasing \mdot. The model predicts that the X-ray
spectra from the two QPO peaks should show similarly different
redshifts as the QPO frequencies. However, this kinematic model has no
way to account for the constant peak separations over a large range in
QPO frequency reported in atoll sources.

In view of the observations of three commensurate frequencies in
several atoll sources (\S1), beat-frequency models (BFMs) are the
mechanism of choice for explaining kHz QPO. The fact that in Sco\,X-1
(this paper), GX\,5$-$1 (van der Klis et al. 1996b) and GX\,17+2 (van
der Klis et al. 1997) we observe HBO and kHz QPO simultaneously shows
conclusively that these can not {\it both} be explained by the
magnetospheric BFM. If this mechanism produces the HBO (\S1), then the
kHz QPO need another model.

Beat-frequency models with the neutron star spin as one of the
participating frequencies predict a constant kHz-peak separation,
which so far was consistent with observations. However, this
prediction is clearly contradicted by our Sco\,X-1 data. In order to
explain the data, such BFMs would have to be modified such that, in
addition to the unseen (in Sco\,X-1) spin frequency $\nu_s$ there is
{\it another} unseen frequency beating with $\nu_s$ to produce one of
the two kHz peaks. The sonic-point beat-frequency model (Miller, Lamb
and Psaltis 1996) may allow such modification (Lamb 1996, priv. comm.). If
this model explains the kHz QPO and the magnetospheric beat-frequency
model the HBO, then in Sco\,X-1 the sonic radius is approximately
$[\nu_u/(\nu_{HBO}+\nu_s)]^{-2/3}$\about0.5 times the magnetospheric radius,
implying the presence of a considerable near-Keplerian flow, where
clumps remain in stable orbit for up to 10$^2$ cycles, well within the
magnetosphere.

\acknowledgments

This work was supported in part by the Netherlands Organization for
Scientific Research (NWO) under grant PGS 78-277 and by the
Netherlands Foundation for Research in Astronomy (ASTRON) under grant
781-76-017. We gratefully acknowledge useful comments on the
manuscript by Stefan Dieters, Erik Kuulkers, Jan van Paradijs and
Walter H.G. Lewin.

\end{document}